# Modeling Interdependent Cybersecurity Threats Using Bayesian Networks: A Case Study on In-Vehicle Infotainment Systems

Sangita Sridar

## Abstract

**Cybersecurity threats are increasingly marked by interdependence, uncertainty, and evolving complexity—challenges that traditional assessment methods such as CVSS, STRIDE, and attack trees fail to adequately capture. This paper reviews the application of Bayesian Networks (BNs) in cybersecurity risk modeling, highlighting their capacity to represent probabilistic dependencies, integrate diverse threat indicators, and support reasoning under uncertainty. A structured case study is presented in which a STRIDE-based attack tree for an automotive In-Vehicle Infotainment (IVI) system is transformed into a Bayesian Network. Logical relationships are encoded using Conditional Probability Tables (CPTs), and threat likelihoods are derived from normalized DREAD scores. The model enables not only probabilistic inference of system compromise likelihood but also supports causal analysis using do-calculus and local sensitivity analysis to identify high-impact vulnerabilities. These analyses provide insight into the most influential nodes within the threat propagation chain, informing targeted mitigation strategies. While demonstrating the potential of BNs for dynamic and context-aware risk assessment, the study also outlines limitations related to scalability, reliance on expert input, static structure assumptions, and limited temporal modeling. The paper concludes by advocating for future enhancements through Dynamic Bayesian Networks, structure learning, and adaptive inference to better support real-time cybersecurity decision-making in complex environments.**

## 1. Introduction: The Importance of Cybersecurity Threat and Risk Assessment

As digital infrastructures grow in complexity and scale, the frequency, severity, and sophistication of cyberattacks have significantly increased. Cybersecurity threats—from ransomware and phishing to supply chain and AI-driven attacks—can severely impact organizational operations, financial stability, and public trust. These threats have escalated in scale and sophistication, posing significant risks to critical infrastructures, enterprise assets, and personal data. As digital transformation advances, the attack surface expands, and threat actors exploit both technical vulnerabilities and human factors.

Cybersecurity risk assessment, therefore, becomes essential in identifying, analyzing, and mitigating potential threats to maintain system confidentiality, integrity, and availability. It supports the triage of vulnerabilities and decision-making around mitigation strategies, enabling organizations to allocate resources efficiently and improve their security posture.

Two key performance metrics in this context are listed below and reducing both metrics is crucial for minimizing the impact of cyber incidents:

- **Mean Time to Identify (MTTI)** – the average time to detect a threat.
- **Mean Time to Contain (MTTC)** – the average time taken to respond and neutralize a threat.

Cybersecurity risk management frameworks typically fall into two categories:

- **Qualitative**: Subjective scoring or ranking of risks, often based on experience or domain knowledge.
- **Quantitative**: Objective, model-based analysis of risks using metrics, statistical tools, or simulations.

Tools such as risk registers, bow-tie models, and attack trees are frequently used to catalog threats and document mitigation strategies. Cybersecurity risk modeling necessitates both the identification of threats (threat modeling) and the quantification of their potential impact (risk assessment).

## Threat Modeling Methodologies

Threat modeling systematically identifies potential vulnerabilities in system architecture before attacks occur. As per NIST SP 800-154, a threat model should support both defensive and offensive evaluations of logical entities, including data flows, hosts, and application boundaries. Some of the methodologies are listed below based on the work by Das et al [19].

**PASTA (Process for Attack Simulation and Threat Analysis)** is a structured, risk-driven methodology developed to align business objectives with security threats in a seven-stage process. It's often used in enterprise environments requiring in-depth analysis of potential attacker behavior and its impact on business operations (UcedaVélez & Morana, 2015 [15]). While it provides a thorough understanding of threats in complex systems, its comprehensive nature makes it resource-intensive, which may not be suitable for small teams or early-stage projects.

**Attack Tree** modeling, introduced by Schneier (1999) [13], represents threats in a tree structure with the attack goal as the root and various possible approaches as branches. It is particularly useful for visualizing the multiple paths an attacker could take. The method is effective for scenario analysis but lacks direct support for threat mitigation and can become unwieldy in complex systems and requires nodes meticulously analyzed and requires employing data flow diagrams for clarity (Shostack, 2014 [14]).

**CVSS (Common Vulnerability Scoring System)** provides standardized metrics for evaluating the severity of software vulnerabilities (Mell, Scarfone, & Romanosky, 2007 [7]). It is commonly used in vulnerability management tools and public vulnerability databases. While CVSS allows for consistent scoring, it lacks contextual modeling and is not sufficient for system-level threat assessment.

**OCTAVE (Operationally Critical Threat, Asset, and Vulnerability Evaluation)** is a risk-based methodology focused on organizational practices and asset protection. Developed at Carnegie Mellon University, it is suitable for evaluating internal risks and supporting strategic planning (Alberts & Dorofee, 2003[2]). However, it is less suited for technical system modeling or identifying fine-grained vulnerabilities in complex environments.

**VAST (Visual, Agile and Simple Threat Modeling)**, developed as part of the ThreatModeler platform, is intended to scale threat modeling practices across enterprise software development pipelines (Agarwal, 2021[1]). It supports both application-level and operational threat models and emphasizes automation. Despite its scalability, the need to model both perspectives can increase complexity and modeling workload.

**LINDDUN** is a privacy-focused methodology that categorizes threats across seven dimensions (e.g., Linkability, Identifiability, Non-repudiation). It is suitable for systems that process sensitive data and aligns with privacy-by-design principles (Deng et al., 2011 [4]). While LINDDUN is effective in highlighting privacy concerns, it lacks automation and is not focused on general cybersecurity threats.

**STRIDE** is a Microsoft-developed model that organizes threats into six categories: Spoofing, Tampering, Repudiation, Information Disclosure, Denial of Service, and Elevation of Privilege (Khan et al., 2016 [6]). It is widely adopted due to its structured threat mapping based on data flow diagrams. Its primary strengths are ease of use, automation support, and applicability to software architecture. However, it may produce variable results depending on model detail and analyst experience.

## 1.2. Risk Assessment Methodologies

Risk assessment evaluates the potential impact and likelihood of identified threats. Some of the commonly used risk assessment is listed below.

**FMVEA (Failure Mode, Vulnerabilities, and Effects Analysis)** combines traditional FMEA with cybersecurity concerns, focusing on identifying and ranking failure modes and vulnerabilities (Schmittner et al., 2014 [12]). It is typically applied in safety-critical domains like automotive and avionics. Its limitation lies in its inability to model chained or multi-stage attacks, making it less ideal for early design phases.

**SHIELD** is a European framework that evaluates systems based on Security, Privacy, and Dependability (SPD) metrics (Macher et al., 2016 [10]). It is effective when detailed system configurations are available and is often used in embedded systems post-implementation. However, it is not well-suited to the early design phase due to its dependency on implementation specifics.

**CHASSIS** integrates safety and security analysis using misuse cases and system behavior modeling. It enables trade-off analysis between safety and security features and is used in early concept modeling (Macher et al., 2016 [10]). Its strength lies in combining formal safety analysis with threat modeling, but the modeling workload can be substantial.

**HEAVENS** (Threat Analysis and Risk Assessment for Embedded Systems) extends STRIDE with automotive-specific considerations, evaluating threats based on attacker capability and impact (Macher et al., 2016 [10]). It's effective in system-level modeling where attacker profiles can be defined. However, its accuracy depends on the availability of detailed architectural data, making it less ideal in early concept stages.

**EVITA** is a framework developed under the European EVITA project to model threats to vehicular IT components using attack trees and functional threat classification (Ruddle et al., 2009 [5]). It supports safety and operational threat evaluation but does not align with ISO 26262 for severity ratings, which limits its regulatory compliance.

**SAHARA** combines the STRIDE model with Hazard Analysis and Risk Assessment (HARA) concepts, using attacker knowledge, required resources, and threat impact to assign security levels (Macher et al., 2015 [11], Macher et al 2016 [10]). It is best applied in the early phases of automotive system development to inform architecture decisions. However, it may miss complex threat chaining or adversarial persistence tactics.

**DREAD** is a lightweight risk scoring method based on five dimensions: Damage, Reproducibility, Exploitability, Affected Users, and Discoverability (Microsoft, 2002 [9]). It is popular in agile software development due to its simplicity. However, its subjective scoring can reduce reliability, and it does not account for contextual risk dependencies.

Traditional risk assessment methodologies often struggle to account for the dynamic and uncertain nature of cyber threats, particularly when historical attack data is limited or incomplete. This limitation underscores the need for probabilistic modeling approaches that can incorporate expert knowledge and uncertainty—of which BNs have emerged as a promising solution.

## 2. The Need for Bayesian Networks in Cybersecurity

BNs overcome these limitations by combining graphical structures with formal probability theory. They enable organizations to shift from static risk catalogues to dynamic, inference-driven threat modeling systems.

Key advantages include:

- **Conditional Dependency Modeling**: BNs use directed acyclic graphs (DAGs) and CPTs to capture how one cyber event probabilistically influences others. This feature is critical in environments where threats are not isolated—such as in cases of lateral movement, privilege escalation, or interdependent system failures (Chockalingam et al., 2022 [18]; Kazeminajafabadi & Imani, 2023 [23]).
- **Real-Time Inference with Incomplete Data**: BNs can compute posterior probabilities for unobserved events (e.g., insider threat or root compromise) using observed evidence (e.g., SIEM alerts). This property is central to security information and event management (SIEM), extended detection and response (XDR), and threat intelligence correlation workflows (Microsoft, 2021[9]).
- **Robustness in Sparse or Noisy Data**: As emphasized by Fenton and Neil (2018) [20], one of the most powerful aspects of BNs is their ability to integrate expert opinion and uncertain data within a coherent probabilistic framework. This is especially important in cybersecurity, where historical breach data is often incomplete, biased, or unavailable.
- **Support for Multi-Stage Attack Modeling**: Extensions like Bayesian Attack Graphs (BAGs) use BN principles to represent how attackers chain exploits across nodes, increasing system-wide compromise probability. These have been successfully applied to network penetration scenarios and ICS/IIoT security environments (Kazeminajafabadi & Imani, 2023 [23]; Poolsappasit et al., 2012 [28]).

Fenton and Neil (2018) [20] also critique overreliance on deterministic models like Monte Carlo simulations or CVSS scores, arguing that such tools break down when systems are complex and interdependent. In contrast, BNs provide a powerful, modular structure for capturing causal and probabilistic relationships, even when input data is uncertain or partially missing.

In summary, BNs offer a mathematically sound and context-aware framework that supports adaptive cybersecurity risk modeling—significantly improving upon traditional registers and scoring systems by integrating threat intelligence, evidence, and uncertainty.

## 3. Bayesian Networks: Mathematical Foundations and Applications in Cybersecurity

BNs are probabilistic graphical models that represent a set of variables and their conditional dependencies using a Directed Acyclic Graph (DAG) (Fenton & Neil, 2018 [20]). Each node corresponds to a variable, while directed edges encode causal or influential relationships. The network's structure supports both qualitative understanding and quantitative analysis.

Bayesian Networks are defined as directed acyclic graphs $G=(V,E)$, where each node $X_i \in V$ is a random variable and each edge $(X_i, X_j) \in E$ represents a conditional dependency. Each node is associated with a Conditional Probability Table (CPT) that defines:

$$P(X_i|Pa(X_i))$$

Where Pa(Xi) are the parent nodes of Xi. The joint probability distribution over all nodes is given by:

$$P(X_1, X_2, ..., X_n) = \prod_{i=1}^{n} P(X_i|Pa(X_i))$$

Mathematically, a BN represents the joint probability distribution over variables X={X1,X2,...,Xn} by factoring it according to its graph structure. This factorization allows efficient inference when new evidence is introduced. For example, if abnormal network activity is observed, the BN can compute posterior probabilities of underlying causes such as malware infection, insider threat, or misconfigured access control (Chockalingam et al., 2022 [18]).

## 4. Applications of Bayesian Networks in Cybersecurity Risk Modelling

BNs have become increasingly prominent in cybersecurity due to their ability to model uncertainty, infer causality, and integrate diverse data sources. Their probabilistic structure allows for reasoning under incomplete or noisy data—conditions typical of real-world cyber environments.

Core Advantages

- **Causal Inference**: BNs support reasoning from observed effects (e.g., anomalous logins) to likely causes (e.g., credential theft), making them particularly valuable in threat hunting, root-cause analysis, and forensic investigations (Fenton & Neil, 2018 [20]; Kazeminajafabadi & Imani, 2023 [23]).
- **Uncertainty Modeling**: BNs are robust in scenarios where historical attack data is sparse or biased, allowing analysts to compute meaningful posterior probabilities using expert knowledge or partial evidence (Fenton & Neil, 2018 [20]).
- **Evidence Fusion**: They enable integration of heterogeneous data sources—such as SIEM alerts, vulnerability scans, and human expert assessments—into a single cohesive risk model, enhancing situational awareness (WRIXTE, 2024 [29]).

These capabilities make BNs well-suited for applications such as:

- **Malware Behavior Prediction**: By modeling observable indicators, BNs have been used to infer the presence of underlying malicious activity (Chockalingam et al., 2022 [18]).
- **Insider Threat Detection**: Incorporating psychosocial and behavioral indicators, BNs have supported the prediction of high-risk insider behavior (Frigault et al., 2008 [21]).
- **Risk-Based Alert Prioritization**: In SIEM and XDR environments, BNs have been employed to reduce alert fatigue by scoring threats based on probabilistic severity and interdependence (Layton, 2024[24]).
- **Adaptive Cyber Defense**: Microsoft's BN-based frameworks have automated attacker profiling, threat tracking, and alert contextualization in enterprise settings (Microsoft, 2021[9]).

**Bayesian Attack Graphs (BAGs)**

An extension of standard BNs, Bayesian Attack Graphs (BAGs) model multistage attacks using system states as nodes and exploit transitions as edges. These models incorporate:

- Exploit probabilities (e.g., from CVSS scores),
- Logical relations such as AND/OR gates for modeling exploit chaining,
- Monitoring imperfections (e.g., false negatives, partial visibility).

Kazeminajafabadi and Imani (2023) [23] demonstrated how BAGs combined with a Minimum Mean Square Error (MMSE) detector significantly improved detection accuracy and reduced false positives in synthetic and simulated networks. Unlike traditional attack graphs, BAGs support dynamic inference, allowing real-time adaptation to new evidence—an essential trait in modern threat landscapes (Poolsappasit et al., 2012 [28]).

**Tooling and Implementation**

Several commercial and open-source tools support BN modeling in cybersecurity:

- **Bayes Server** offers a GUI-based platform for building BNs, conducting sensitivity analysis, and simulating risk scenarios.
- **Python libraries** such as pgmpy and pyAgrum enable programmatic BN construction and inference, suitable for integrating into security analytics pipelines.

Layton (2024) [24] highlights how BNs are increasingly used by small and medium businesses (SMBs) and healthcare providers due to their transparency, modularity, and lower data demands compared to opaque machine learning models.

**Cross-Domain Applications**

Bayesian Networks have been effectively applied across a variety of cybersecurity domains:

- **Network Risk Assessment**: Frigault et al. (2008) [21] used Dynamic Bayesian Networks (DBNs) to assess the risk across interconnected network services, capturing dynamic interactions and vulnerabilities under evolving conditions.
- **Intrusion Detection and Adaptive Monitoring**: MMSE-enhanced BAGs (Kazeminajafabadi & Imani, 2023 [23]) enabled optimal placement of sensors and real-time adjustment of detection policies.
- **Cyber Risk Quantification**: Babatunde et al. (2024) [17] demonstrated how BNs could guide decision-making by simulating attack scenarios and assessing their business impact.
- **Industrial IoT (IIoT)**: Karim et al. (2024) [22] developed a hybrid BN–swarm intelligence model for IIoT threat detection, using feature selection and evolutionary optimization to improve accuracy.
- **Real-Time Threat Assessment**: Pappaterra et al. (2020) [26] embedded BNs in the DETECT framework for live cybersecurity monitoring, enabling systems to update risk assessments in response to incoming data.

# 5. Case Demonstration: Modeling Interdependent STRIDE Threats Using Bayesian Networks in an IVI System

The case study in this paper draws upon a detailed STRIDE-based threat modeling framework applied to a modern In-Vehicle Infotainment (IVI) system, originally presented by Das et al. (2024) [19]. The IVI system under consideration is a highly interconnected, cyber-physical subsystem embedded in automotive environments. It integrates a diverse range of components enabling real-time media, communication, and vehicle information services.

The system architecture comprises:

- A central onboard computer (OBC) acting as the processing hub.
- Interfaces for Bluetooth, Wi-Fi and NFC, facilitating external communication.
- A CAN bus network used to interact with internal Electronic Control Units (ECUs).
- Additional peripheral elements such as a touchscreen interface, rear screens, audio system, GPS, USB ports, digital radio, camera, and various car automation sensors (e.g., parking assistant, temperature sensor).

This IVI system supports bidirectional communication between external devices (e.g., smartphones, cloud services) and internal vehicle subsystems. While enhancing driver convenience and connectivity, this heterogeneity introduces critical cybersecurity vulnerabilities, especially across trust boundaries—such as data flows between Bluetooth/Wi-Fi modules and the onboard computer, or between the onboard computer and CAN bus.

To assess these risks, Das et al. conducted a component-level STRIDE threat modeling exercise, identifying 34 distinct threats across interactions such as Bluetooth_to_OBC, WiFi_to_OBC, and CAN_to_OBC. Each threat was further evaluated using SAHARA and DREAD methodologies, producing detailed risk ratings and prioritization. These findings provide a rich, structured threat ontology which this paper maps into a BN framework to enable probabilistic inference, threat propagation modeling, and dynamic risk evaluation.

In traditional STRIDE-based models, threats are often mapped in isolation without considering their probabilistic relationships or downstream implications. However, real-world attacks often unfold as interconnected sequences. A BN enables modeling these relationships explicitly, supporting probabilistic inference, evidence fusion, and risk propagation analysis.

The Figure 1 below illustrates an example model Bayesian Network constructed from STRIDE threat categories affecting an IVI subsystem as adapted from Das et al (2024), where a tampering event over Bluetooth initiates a cascade of other threats:

*Figure 1: Bayesian Network of STRIDE Threats in an IVI System (adapted from Das et al., 2024)*

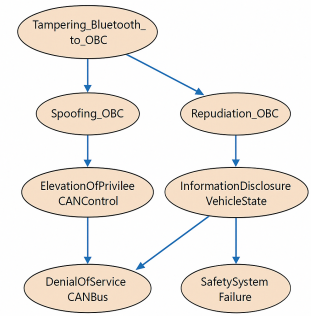

## 5.1 Methodology: Bayesian Network Construction for STRIDE Threat Propagation in IVI Systems

This paper focuses on foundational BN-based modeling, with an understanding that more sophisticated techniques like

DBNs and structure learning exist for future extensions. To evaluate the effectiveness of BNs in modeling interdependent cybersecurity threats, a structured methodology that transforms a STRIDE-based attack tree into a probabilistic graphical model was applied in this case study. This approach supports inference over cascading threats within an In-Vehicle Infotainment (IVI) system, enabling probabilistic reasoning, dynamic threat propagation analysis, and risk-aware decision-making under uncertainty.

*5.1.1 Model-to-Model Transformation*

This paper adopted the model-to-model transformation process outlined by Pappaterra and Flammini (2018) [27] and Hachem [31] et al work, wherein structured attack trees (ATs) are systematically mapped into BN elements. Each node in the attack tree, derived from Table 7 of Das et al. (2024), represents a threat instance using following schema:

[STRIDE] _[Component Interaction] _[Threat Number]_[Two Word Threat Description]

Where,

- STRIDE is denoted by first letter of the threat type.
- Component Interaction, and Threat Number uses the same format as Table 7 of Das et al. (2024) [19] paper such as WiFitoOBC or BluetoothtoOBC.
- Two Word Threat Description is the summary of the threat description as described in Table 7 of Das et al. (2024) [19] paper.

The transformation process consisted of the following rules:

- Leaf nodes in the attack tree were converted into root nodes in the BN, each representing a STRIDE-identified threat.
- Intermediate AT nodes, composed of logical gates (AND/OR), were transformed into conditional BN nodes, where dependencies were encoded through directed edges.
- Operation logic was preserved by defining node-specific CPTs to reflect the combinatorial influence of parent nodes.

The attack tree begins with the ultimate adversarial goal "Safety Critical System Compromise", representing failure of safety-critical vehicle operations. This outcome can be reached via three main attack paths:

- **Initial Recon/Entry Path (OR Logic):** Begins with reconnaissance attacks like Data Sniffing over Bluetooth or Credential Theft via Wi-Fi.
    - I_BluetoothtoOBC_25_DataSniffing is triggered by T_BluetoothtoOBC_26_UnauthorizedControl or D_BluetoothtoOBC_24_OverloadAttack.
    - I_WiFitoOBC_17_CredentialTheft is triggered by D_WiFitoOBC_16_ServiceDenial or T_WiFitoOBC_18_DataAlteration

- **CAN Control Path (AND Logic):** Requires successful exploitation of:
    - I_CBtoOBC_33_InfoSniffing by sniffing the data flow on the CAN bus
    - T_CBtoOBC_34_MessageAlteration (tampering with messages) **and**
    - S_OBC_1_ProcessImpersonation (mimicking valid processes), the latter being triggered by R_OBC_3_MaliciousExploitation **or** T_OBC_2_CommandTampering

- **Disruption of Vehicle Functionality (OR Logic):** Involves either:
    - T_OBC_2_CommandTampering
    - Or service disruption through D_OBC_5_ServiceDisruption, which itself depends on I_OBC_4_PrivacyBreach or E_OBC_6_PrivilegedOperations

The attack tree uses OR and AND gates to represent the logical combinations needed to escalate through each threat path, culminating in system compromise (Figure 2)

*Figure 2: Attack tree developed with the adversarial goal "Safety Critical System Compromise" for the IVI System described in Das et al., 2024 paper*

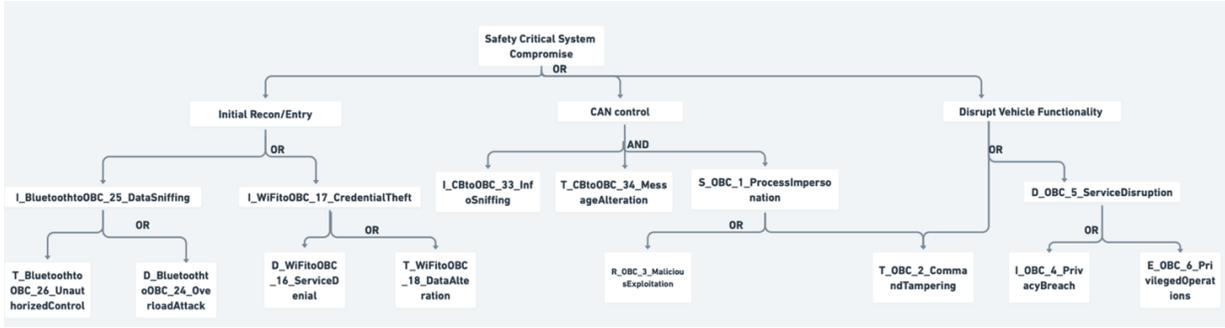

BN was derived based on the model-to-model transformation (Figure 3). Each node corresponds to a threat, and directed edges represent causal or conditional influence.

- Root Nodes (e.g., T_BluetoothtoOBC_26_UnauthorizedControl) have assigned prior probabilities based on DREAD scores.
- Intermediate Nodes like I_BluetoothtoOBC_25_DataSniffing or I_WiFitoOBC_17_CredentialTheft are activated conditionally, influenced by child threats using OR logic.
- Key conditional nodes like CAN_Control are activated only when multiple threats (e.g., message alteration AND impersonation) occur simultaneously.
- Disruption and compromise paths converge at Safety_Critical_System_Compromise, the final output node.

What makes the BN powerful is its ability to:

- Perform probabilistic inference (e.g., "If onboard computer alteration is detected, how likely is CAN control compromise?")
- Handle uncertainty and interdependence, allowing real-time system risk evaluation based on partial observations.

*Figure 3: Bayesian network developed with the adversarial goal "Safety Critical System Compromise" for the IVI System described in Das et al., 2024 paper based on the model-to-model transformation as described by Pappaterra and Flammini (2018)*

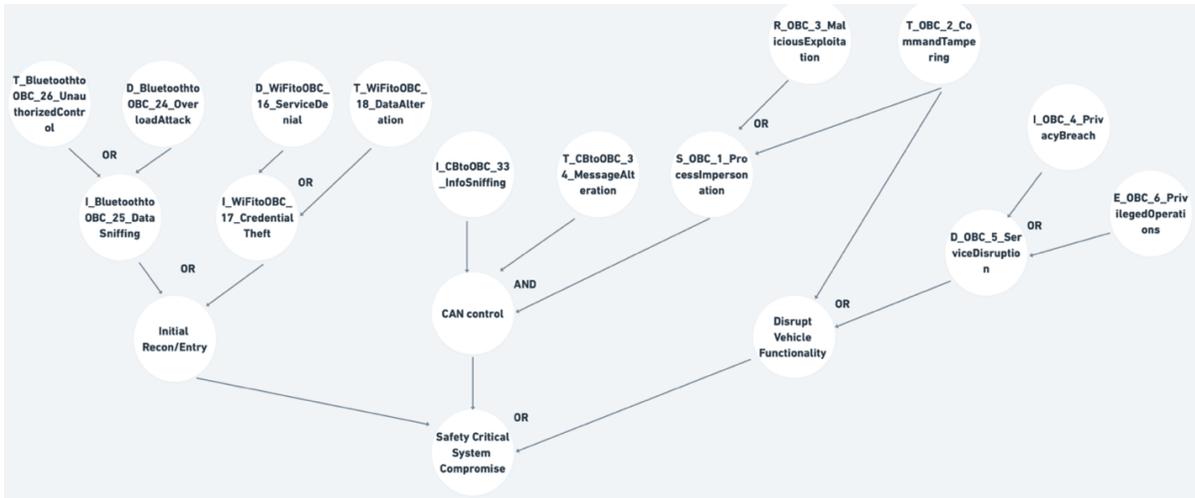

### 5.1.2 Data Population of CPTs

Following the frameworks established by Pappaterra [26] and Hachem et al. [31], the Conditional Probability Tables (CPTs) of the Bayesian Network (BN) were populated using a hybrid data-driven approach that incorporated structured threat evaluation metrics from the Das et al. study. In particular, prior probabilities for the root nodes (e.g., T_BluetoothtoOBC_26_UnauthorizedControl) were derived from DREAD scores, focusing on three key components: Exploitability (E), Discoverability (D), and Reproducibility (R). These scores were normalized to a [0,1] scale using the following procedure:

1. **Normalization**: Each raw DREAD component score (on a 1–3 scale) was divided by 3 to bring it into the [0,1] range.

2. **Weighted Aggregation**: Component weights were assigned based on their relative importance—Exploitability (50%), Discoverability (30%), and Reproducibility (20%).

3. **Likelihood Calculation**: A weighted likelihood was computed using the formula:
   **Likelihood = (0.5 × E) + (0.3 × D) + (0.2 × R)**

This calculation ensured that each root node's prior probability reflected a meaningful combination of severity and exploitability.

Once the root node priors were assigned, marginal probabilities for all intermediate and terminal nodes were calculated using a bottom-up propagation algorithm that incorporates both logical structure and cyber threat intelligence. Specifically, CPTs for child nodes were assigned grounded in attacker behavior patterns defined in the MITRE ATT&CK framework:

1. Nodes were processed hierarchically, starting from root nodes and moving through the network based on dependency structure.

2. For nodes with parents, the law of total probability was applied to integrate all possible parent state combinations using corresponding CPT entries.

3. In multi-parent configurations, joint probabilities of all parent node states were calculated and multiplied by the corresponding row in the node's MITRE-informed CPT.

4. Computed probabilities were recursively propagated forward, treating each evaluated node as evidence for its descendants.

5. Final marginal probabilities were verified to ensure all distributions summed to 1.

The BN model was implemented in Python using the pyAgrum library. The modeling process involved the following key steps:

- **Node Initialization**: Twenty binary variables were instantiated using gum.LabelizedVariable, each representing a vulnerability, system behavior, or adversarial event, with two states: 0 (False) and 1 (True).

- **Topology Definition**: Directed edges were added between nodes to encode causal relationships and threat propagation logic, using both OR and AND semantics to accurately represent attack dynamics.

- **CPT Assignment**: Root nodes such as T_26 and T_18 received prior probabilities via .fillWith(), informed by expert assessments and CVSS-style scoring. Intermediate and final nodes (e.g., I_17, CAN_Control, SystemCompromise) were manually configured with conditional probabilities based on logical gate semantics and expert-derived estimates.

Inference was conducted using the gum.LazyPropagation algorithm to compute marginal probabilities across the network in the absence of observed evidence. This represented the system's baseline risk posture. Results were visualized using pyagrum.lib.notebook.

This combined methodology provided a transparent and consistent framework for quantitative cybersecurity risk assessment. It captured both the individual threat likelihoods and their complex interdependencies, ultimately enabling the computation of the system-wide compromise probability.

The calculated prior probabilities (PPTs) for the root nodes are summarized in the following Table 1

*Table 1: The table shows the root nodes and its corresponding DREAD score and prior probabilities*

| Node | Threat No. | Reproducibility (R) | Exploitability (E) | Discoverability (D) | Prior Probabilities (PPT) |
|---|---|---|---|---|---|
| )T_BluetoothtoOBC_26_UnauthorizedControl | 26 | 2 | 2 | 2 | [False: 0.33, True: 0.67] |
| D_BluetoothtoOBC_24_OverloadAttack | 24 | 3 | 2 | 2 | [False: 0.27, True: 0.73] |
| D_WiFitoOC_16_ServiceDenial | 16 | 3 | 2 | 2 | [False: 0.27, True: 0.73] |
| T_WiFitoOBC_18_DataAlteration | 18 | 2 | 3 | 2 | [False: 0.17, True: 0.83] |
| I_CBtoOBC_33_InfoSniffing | 33 | 2 | 2 | 2 | [False: 0.33, True: 0.67] |
| T_CBtoOBC_34_MessageAlteration | 34 | 2 | 2 | 2 | [False: 0.33, True: 0.67] |
| R_OBC_3_MaliciousExploitation | 3 | 2 | 2 | 2 | [False: 0.33, True: 0.67] |
| T_OBC_2_CommandTampering | 2 | 2 | 2 | 2 | [False: 0.33, True: 0.67] |
| I_OBC_4_PrivacyBreach | 4 | 2 | 2 | 2 | [False: 0.33, True: 0.67] |
| E_OBC_6_PrivilegedOperations | 6 | 2 | 2 | 3 | [False: 0.23, True: 0.77] |

Conditional probabilities for intermediate nodes were designed to reflect logical relationships using OR and AND gate semantics. For nodes governed by OR logic, the CPTs were configured to produce a high probability output if at least one of the parent nodes was in the active (True) state. Conversely, for nodes modeled with AND logic, the CPTs were calibrated such that a high output probability would occur only when all parent nodes were simultaneously active. This approach ensured that the probabilistic behavior of the network accurately mirrored logical dependencies between contributing threats.

The probability of the node *I_BluetoothtoOBC_25_DataSniffing* was calculated based on two parent threats: *T_BluetoothtoOBC_26_UnauthorizedControl* and *D_BluetoothtoOBC_24_OverloadAttack (Table 2)*. The CPT followed an OR-logic structure. Data sniffing becomes increasingly likely when either unauthorized Bluetooth control or traffic overload occurs. Unauthorized control (T1546) directly enables access to system resources or services that may emit sensitive information, while overload attacks (T1499) can create conditions where data leakage or reduced encryption safeguards occur. The CPT reflects this: risk is low (0.20) in absence of threats, moderate (0.60) with overload alone, high (0.80) with unauthorized control alone, and near certain (0.90) when both threats are active. By applying the law of total probability and multiplying each parent state combination by its corresponding CPT entry, the marginal probability of *I_BluetoothtoOBC_25_DataSniffing* was computed to be approximately **74.7%**. This result captures both the direct and indirect impacts of Bluetooth tampering and system overload on the likelihood of data interception.

*Table 2: The table shows CPT for I_BluetoothtoOBC_25_DataSniffing node based on OR Logic between T_BluetoothtoOBC_26_UnauthorizedControl and D_BluetoothtoOBC_24_OverloadAttack nodes*

| T_BluetoothtoOBC_26_UnauthorizedControl | D_BluetoothtoOBC_24_OverloadAttack | I_BluetoothtoOBC_25_DataSniffing 0 | I_BluetoothtoOBC_25_DataSniffing 1 |
|---|---|---|---|
| 0 | 0 | 0.8000 | 0.2000 |
| 0 | 1 | 0.4000 | 0.6000 |
| 1 | 0 | 0.2000 | 0.8000 |
| 1 | 1 | 0.1000 | 0.9000 |

The probability of the node I_WiFitoOBC_17_CredentialTheft was determined based on two parent threats: T_WiFitoOBC_18_DataAlteration and D_WiFitoOBC_16_ServiceDenial (*Table 3*). The CPT followed an OR-logic structure. Credential theft risk grows substantially when attackers manipulate transmitted data or cause Wi-Fi-based disruptions. Data alteration (T1565.001) is a strong enabler for spoofing login packets, and DoS attacks (T1498) can expose data by forcing the system into less secure states. The CPT reflects this: low baseline risk (0.08), modest risk (0.25) from DoS alone, high risk (0.70) from data manipulation alone, and very high risk (0.90) when both are active. Using the law of total probability, the marginal probability of I_WiFitoOBC_17_CredentialTheft was computed to be approximately **73.7%**. This result reflects the elevated risks associated with both active tampering of Wi-Fi or cellular data flows and service disruption attacks.

*Table 3: The table shows CPT for I_WiFitoOBC_17_CredentialTheft node based on OR Logic between T_WiFitoOBC_18_DataAlteration and D_WiFitoOBC_16_ServiceDenial nodes*

| D_WiFitoOBC_16_ServiceDenial | T_WiFitoOBC_18_DataAlteration | I_WiFitoOBC_17_CredentialTheft 0 | I_WiFitoOBC_17_CredentialTheft 1 |
|---|---|---|---|
| 0 | 0 | 0.9200 | 0.0800 |
| 0 | 1 | 0.3000 | 0.7000 |
| 1 | 0 | 0.7500 | 0.2500 |
| 1 | 1 | 0.1000 | 0.9000 |

The probability of the node S_OBC_1_ProcessImpersonation was assessed based on two parent threats: R_OBC_3_MaliciousExploitation and T_OBC_2_CommandTampering (Table 4*). The CPT followed an OR-logic structure. Process impersonation stems from either software exploitation or direct manipulation of control commands. T1203 provides the attacker with execution capability needed to inject or spoof processes, while T1565 allows crafted commands to mimic valid behaviors. The CPT reflects high risk (0.85) for either parent alone and near certainty (0.98) when both are active. With neither, impersonation is unlikely (0.10). Applying the law of total probability across all parent state combinations, the marginal probability of *S_OBC_1_ProcessImpersonation* was computed to be approximately **82.7%**.

*Table 4: The table shows CPT for S_OBC_1_ProcessImpersonation node based on OR Logic between R_OBC_3_MaliciousExploitation and T_OBC_2_CommandTampering nodes*

| R_OBC_3_MaliciousExploitation | T_OBC_2_CommandTampering | S_OBC_1_ProcessImpersonation 0 | S_OBC_1_ProcessImpersonation 1 |
|---|---|---|---|
| 0 | 0 | 0.9000 | 0.1000 |
| 0 | 1 | 0.1500 | 0.8500 |
| 1 | 0 | 0.1500 | 0.8500 |
| 1 | 1 | 0.0200 | 0.9800 |

The probability of the node *D_OBC_5_ServiceDisruption* was determined based on two parent threats: *E_OBC_6_PrivilegedOperations* and *I_OBC_4_PrivacyBreach (*Table 5*)*. The CPT followed an OR-logic structure. Service disruption becomes highly probable when attackers gain access to privileged accounts or internal sensitive data. T1078 directly enables unauthorized system manipulation, while T1530 informs tailored DoS attempts. Alone, each parent leads to significant risk (0.60–0.85), and together they escalate the likelihood to 0.95. A low baseline risk (0.20) reflects default system resilience. Applying the law of total probability, the marginal probability of *D_OBC_5_ServiceDisruption* was computed to be approximately **81.4%**. This result reflects the cumulative impact of access control failures and data privacy breaches

*Table 5: The table shows CPT for D_OBC_5_ServiceDisruption node based on OR Logic between E_OBC_6_PrivilegedOperations and I_OBC_4_PrivacyBreach nodes*

| E_OBC_6_PrivilegedOperations | I_OBC_4_PrivacyBreach | D_OBC_5_ServiceDisruption 0 | D_OBC_5_ServiceDisruption 1 |
|---|---|---|---|
| 0 | 0 | 0.8000 | 0.2000 |
| 0 | 1 | 0.4000 | 0.6000 |
| 1 | 0 | 0.1500 | 0.8500 |
| 1 | 1 | 0.0500 | 0.9500 |

The probability of the node *Initial_Recon* was evaluated based on two parent
threats: *I_BluetoothtoOBC_25_DataSniffing* and *I_WiFitoOBC_17_CredentialTheft* (Table 6). The CPT follows an OR-logic structure. Initial Recon represents the adversary's first foothold. Either Bluetooth sniffing (T1040) or credential theft (T1555) is sufficient to begin reconnaissance within the system. The CPT shows that the presence of either alone result in a high likelihood (0.75), and their combination elevates it to near certainty (0.95). With neither active, the threat is minimal (0.10), reflecting the need for a reconnaissance vector to enable further action. Applying the law of total probability across all parent combinations, the marginal probability of Initial_Recon was computed to be approximately **81.7%.** This reflects the heightened risk associated with multiple attack vectors and supports a robust quantitative analysis of initial access vulnerabilities in automotive cybersecurity modeling

*Table 6: The table shows CPT for Initial_Recon node based on OR Logic between I_BluetoothtoOBC_25_DataSniffing and I_WiFitoOBC_17_CredentialTheft nodes*

| I_WiFitoOBC_17_CredentialTheft | I_BluetoothtoOBC_25_DataSniffing | Initial_Recon 0 | Initial_Recon 1 |
|---|---|---|---|
| 0 | 0 | 0.9000 | 0.1000 |
| 0 | 1 | 0.2500 | 0.7500 |
| 1 | 0 | 0.2500 | 0.7500 |
| 1 | 1 | 0.0500 | 0.9500 |

The CPT for the CAN_control threat models the likelihood of an adversary gaining full control over the CAN bus by combining three parent threats: I_CBtoOBC_33_InfoSniffing, T_CBtoOBC_34_MessageAlteration, and S_OBC_1_ProcessImpersonation (Table 7). Info sniffing provides critical reconnaissance, message alteration enables direct tampering with vehicle operations, and process impersonation allows adversaries to bypass security by mimicking legitimate processes. CAN bus control requires layered attacks: sniffing to learn protocol (T1040), tampering with message data (T1565.002), and issuing spoofed instructions from a legitimate-looking process (T1055.012). The CPT models an AND-like structure with probabilistic weighting: single-parent activation leads to low/moderate risk (0.15–0.30), dual combinations lead to mid-to-high risk (0.50–0.70), and all three threats together push risk to near certainty (0.95). Minimal risk (0.05) is assigned when none are active. This CPT structure captures the layered, multi-stage progression typical of real-world CAN bus attacks and emphasizes the necessity of defense-in-depth strategies in automotive cybersecurity. The probability assignments are grounded in expert judgment and form a quantitative basis for evaluating CAN bus security vulnerabilities. Applying the law of total probability across all parent combinations, the marginal probability of CAN_control was computed to be approximately **64.4%.**

*Table 7: The table shows CPT for CAN_control node based on AND Logic between I_CBtoOBC_33_InfoSniffing, T_CBtoOBC_34_MessageAlteration, and S_OBC_1_ProcessImpersonation nodes*

| I_CBtoOBC_33_InfoSniffing | S_OBC_1_ProcessImpersonation | T_CBtoOBC_34_MessageAlteration | CAN_Control 0 | CAN_Control 1 |
|---|---|---|---|---|
| 0 | 0 | 0 | 0.9500 | 0.0500 |
| 0 | 0 | 1 | 0.7000 | 0.3000 |
| 0 | 1 | 0 | 0.8000 | 0.2000 |
| 0 | 1 | 1 | 0.4000 | 0.6000 |
| 1 | 0 | 0 | 0.8500 | 0.1500 |
| 1 | 0 | 1 | 0.3000 | 0.7000 |
| 1 | 1 | 0 | 0.5000 | 0.5000 |
| 1 | 1 | 1 | 0.0500 | 0.9500 |

The probability of the node Disrupt_Vehicle_Functionality was assessed based on two parent threats: T_OBC_2_CommandTampering and D_OBC_5_ServiceDisruption (*Table 8*). The CPT follows an OR-logic structure. Vehicle function disruption results when control commands are corrupted, or the system is overwhelmed. T1565 enables precise tampering, while T1499 affects availability. Each parent causes high risk alone (0.75–0.80), and together they push the probability to 0.95. A low risk (0.10) is assigned in absence of both. After calculating the marginal probability using the law of total probability, Disrupt_Vehicle_Functionality was determined to have a likelihood of approximately 83.3%. This outcome highlights the critical vulnerabilities posed by denial-of-service conditions and unauthorized command injection within vehicle control systems.

*Table 8: The table shows CPT for Disrupt_Vehicle_Functionality node based on OR Logic between T_OBC_2_CommandTampering and D_OBC_5_ServiceDisruption nodes*

| D_OBC_5_ServiceDisruption | T_OBC_2_CommandTampering | Disrupt_Vehicle_Functionality 0 | Disrupt_Vehicle_Functionality 1 |
|---|---|---|---|
| 0 | 0 | 0.9000 | 0.1000 |
| 0 | 1 | 0.2500 | 0.7500 |
| 1 | 0 | 0.2000 | 0.8000 |
| 1 | 1 | 0.0500 | 0.9500 |

The CPT for the final System Compromise Node, Safety_Critical_System_Compromise, reflects the logical OR relationship among its three parent nodes: Initial_Recon/Entry, CAN_Control, and Disrupt_Vehicle_Functionality (*Table 9*). This design assumes that a successful attack along any one of these major threat paths is sufficient to substantially raise the likelihood of systemic compromise, while the presence of multiple active paths further amplifies the risk.

To ensure consistency with Bayesian Network requirements, each row in the CPT represents a unique combination of parent states with a corresponding probability distribution that sums to 1. The probability assignments are grounded in domain knowledge from cybersecurity literature and normalized DREAD scores that indicate high threat severity and exploitability. Specifically, a low residual risk (approximately 1%) is assigned when none of the parent threats are active, reflecting background uncertainty in cyber-physical systems. When at least one parent threat is active, the probability of compromise increases substantially (typically ranging from 75% to 95%), and when all three threats are present simultaneously, the risk approaches near certainty (~99%).

Following OR-gate computation, the marginal probability of Safety_Critical_System_Compromise was calculated as approximately 93.5%, highlighting the nonlinear escalation of system risk in interdependent threat landscapes such as those modeled under STRIDE-based frameworks. This configuration adheres to the semantics of probabilistic inference while accurately capturing the compounding effects of multiple concurrent adversarial actions on vehicle system safety

*Table 9: The table shows CPT for final node Safety_Critical_System_Compromise*

| CAN_Control | Initial_Recon | Disrupt_Vehicle_Functionality | SystemCompromise 0 | SystemCompromise 1 |
|---|---|---|---|---|
| 0 | 0 | 0 | 0.9500 | 0.0500 |
| 0 | 0 | 1 | 0.2000 | 0.8000 |
| 0 | 1 | 0 | 0.2500 | 0.7500 |
| 0 | 1 | 1 | 0.0500 | 0.9500 |
| 1 | 0 | 0 | 0.2500 | 0.7500 |
| 1 | 0 | 1 | 0.0500 | 0.9500 |
| 1 | 1 | 0 | 0.0500 | 0.9500 |
| 1 | 1 | 1 | 0.0100 | 0.9900 |

## 5.2 Model Interpretation: Safety Compromise Risk in an IVI System

The constructed BN models the probabilistic propagation of cyber threats culminating in a SystemCompromise event within an In-Vehicle Infotainment (IVI) system. It incorporates both root-level vulnerabilities (e.g., T_BluetoothtoOBC_26_UnauthorizedControl, D_WiFitoOBC_16_ServiceDenial) and higher-order composite threats (e.g., CAN_Control, Disrupt_Vehicle_Functionality), enabling interpretable, real-time risk assessment under uncertainty.

The BN captures inference across three principal threat propagation paths:

- Initial Reconnaissance – Represents early-stage attack vectors such as data sniffing and credential theft.
- CAN_Control – Reflects unauthorized manipulation of vehicle control networks, such as CAN bus spoofing or message alteration.
- Disrupt_Vehicle_Functionality – Models attacks that degrade or disable core vehicle operations via component-level exploitation.

Based on the methodology described in Section 5.1, the model computes the following posterior probabilities (Figure 4):

- P(I_BluetoothtoOBC_25_DataSniffing = True) ≈ 76.73%
- P(Initial_Recon = True) ≈ 81.69%
- P(CAN_Control = True) ≈ 64.43%

**Baseline Observations**

In the absence of observed attacks, the system estimates a low baseline risk of compromise: P(SystemCompromise=True) ≈ 0.0652, confirming that an uncompromised system operates within a low-risk envelope. However, the model demonstrates nonlinear risk escalation when upstream threats are activated, due to the dependencies encoded in CPTs.

**Threat Chain Behavior**

When T_BluetoothtoOBC_26_UnauthorizedControl is activated (prior: 67%), the posterior probability of I_BluetoothtoOBC_25_DataSniffing increases to 76.73%, which then propagates forward to Initial_Recon = 81.69%. Similarly, T_WiFitoOBC_18_DataAlteration (83%) and D_WiFitoOBC_16_ServiceDenial (73%) jointly raise I_WiFitoOBC_17_CredentialTheft to 73.69% and also contribute to the elevation of Initial_Recon.

These contribute to an increased belief in SystemCompromise, demonstrating how even partial threat chain activation significantly alters system risk.

The CAN_Control node—an AND-composed node—requires joint activation of I_CBtoOBC_33_InfoSniffing, T_CBtoOBC_34_MessageAlteration, and S_OBC_1_ProcessImpersonation. With all three nodes active at ~67% or higher, CAN_Control=True is inferred at 64.43%. In the third path, Disrupt_Vehicle_Functionality=True is inferred at 83.25%, driven by high prior likelihoods of D_OBC_5_ServiceDisruption (81.37%) and its parents.

Collectively, these threat vectors result in a posterior compromise probability of 93.48%, indicating a near-certain risk in realistic multi-vector attack scenarios.

**Key Insights**

**Propagation Through Alternative Paths**
The Initial_Recon path independently escalates risk without CAN pathway involvement, demonstrating how entry-level attacks (e.g., data sniffing or credential theft) can be just as damaging if unaddressed.

**Nonlinear Risk Escalation**
The model reflects that multiple moderate threats can cumulatively drive high system risk. For instance, three ~70% activated sub-threats cause CAN_Control=True inference to jump to ~64.4%.

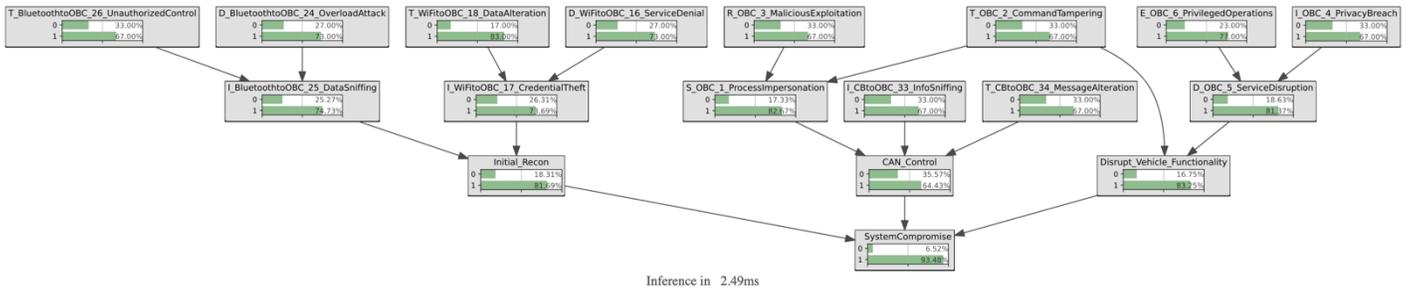

*Figure 4: Inference of the IVI system*

**Logical Gate Fidelity**

Unlike flat scoring models, the BN respects the AND/OR logic of threat interdependence. This ensures greater fidelity in simulating realistic multi-phase attack strategies, such as requiring simultaneous spoofing and tampering to impact CAN bus operations.

**Prioritization of Mitigation Strategies**

The model supports prioritizing high-impact vulnerabilities. For example, reducing the attack probability on CredentialTheft or MessageAlteration has a disproportionately strong downstream effect on compromise risk, making them valuable mitigation targets.

# 6. Causal Analysis and Intervention Impact

To move beyond correlation and quantify the true influence of individual threats on system-level risk, the paper demonstrates **causal impact analysis** using a *do-calculus* approach on the BN. While traditional BN inference computes posterior probabilities based on observations (i.e., P(Y|X)), causal reasoning estimates the effect of *actively forcing a node into a specific state*, expressed as P(SystemCompromise | do(X=1)).

In practical terms, this means simulating what happens when a threat is forcibly activated (e.g., a Bluetooth-based attack occurs), *regardless of its upstream dependencies*. This distinction is critical in cybersecurity: not all high-probability nodes have high causal influence, and not all influential threats are obvious from their marginal probability alone.

**Causal vs. Observational Reasoning**

- The table below compares posterior probabilities (inferred from natural propagation in the BN) with interventional probabilities (computed via causal impact analysis using do(X=1)) for both root threats and intermediate nodes (Table 10).

**Interpretation of Results**

- Most nodes showed a significantly higher impact under causal intervention than what their natural BN propagation suggested.

- CAN_Control had the greatest causal influence on SystemCompromise, increasing risk by over 32% when forcibly activated—despite a relatively modest posterior probability of 64.43%. This reflects its central role in consolidating multiple threat vectors.

- Several upstream nodes, like T_2_CommandTampering and T_34_MessageAlteration, showed high causal effects (>27%), demonstrating that interventions on seemingly modest threats can have system-wide consequences.

*Table 10: The table shows casual probabilities for root and middle nodes*

| Node | Posterior P(Node=1) | P(SystemCompromise \| do(Node=1)) | Δ Increase in Risk |
|---|---|---|---|
| T_26_UnauthorizedControl | 67.00% | 94.00% | +27.00% |
| D_24_OverloadAttack | 73.00% | 93.70% | +20.70% |
| T_18_DataAlteration | 83.00% | 93.92% | +10.92% |
| D_16_ServiceDenial | 73.00% | 93.69% | +20.69% |
| R_3_MaliciousExploitation | 67.00% | 93.84% | +26.84% |
| T_2_CommandTampering | 67.00% | 95.05% | +28.05% |
| I_33_InfoSniffing | 67.00% | 94.54% | +27.54% |
| T_34_MessageAlteration | 67.00% | 94.90% | +27.90% |
| E_6_PrivilegedOperations | 77.00% | 94.07% | +17.07% |
| I_4_PrivacyBreach | 67.00% | 93.80% | +26.80% |
| I_25_DataSniffing | 76.73% | 94.53% | +17.80% |
| I_17_CredentialTheft | 73.69% | 94.56% | +20.87% |
| Initial_Recon | 81.69% | 95.88% | +14.19% |
| S_1_ProcessImpersonation | 82.67% | 94.06% | +11.39% |
| CAN_Control | 64.43% | 97.11% | +32.68% |
| D_5_ServiceDisruption | 81.37% | 94.55% | +13.18% |
| Disrupt_Vehicle_Functionality | 83.25% | 96.13% | +12.88% |

- Intermediate nodes like Initial_Recon and Disrupt_Vehicle_Functionality also showed high sensitivity to intervention, validating their role as critical pivot points in attack propagation.

**Security Implications**

- Prioritization: Causal analysis enables rational prioritization of threats for mitigation. Focusing on high-causal-impact nodes (like CAN control and Command Tampering) delivers maximum risk reduction.
- Defense Strategy: Nodes with high interventional impact should be monitored in real-time and subjected to the strongest access control, anomaly detection, and hardening strategies.
- Audit & Compliance: These insights provide a defensible basis for targeting specific controls in risk registers and security audits.

# 7. Sensitivity Analysis of Attack Propagation in the Bayesian Network

To assess the influence of individual nodes on the system-level security outcome, the paper demonstrates local sensitivity analysis using the constructed BN, which models cyberattack propagation across interconnected system components (based on Hachem et al work). This process evaluates the robustness of the output probability of SystemCompromise=True against perturbations in the input parameters.

The sensitivity analysis procedure consisted of four main steps. First, a **baseline computation** was conducted using probabilistic inference via the LazyPropagation algorithm in PyAgrum to calculate the baseline probability of the top-level node, *SystemCompromise=True*, which signifies a successful cyberattack within the modeled system. Next, in the **PPT perturbation phase**, each root node—representing individual vulnerabilities without parent dependencies—had its "True" state probability increased by 0.1 (with an upper limit of 0.99). The modified Bayesian Network was then used to recompute *P(SystemCompromise=True)*, and the change from the baseline was recorded to assess the marginal influence of each root node. In the **CPT perturbation phase**, targeting intermediate nodes (such as *Initial_Recon* or *CAN_Control*), every combination of parent states was iterated through, with each associated CPT entry increased by 0.1 while remaining bounded within [0.01, 0.99]. The resulting changes in *P(SystemCompromise=True)* were recorded, and the average of these differences was calculated to yield a sensitivity score for each intermediate node. Finally, during the **sensitivity aggregation and visualization step**, sensitivity scores for both root and intermediate nodes were compiled into a dictionary. A bar chart (

*Figure 5*) was then generated using Matplotlib to visualize the relative impact of each node's perturbation on the system compromise probability, with nodes sorted and labeled according to their role in the Bayesian Network.

*Figure 5: Sensitivity analysis for the IVI system based on Hachem et al work*

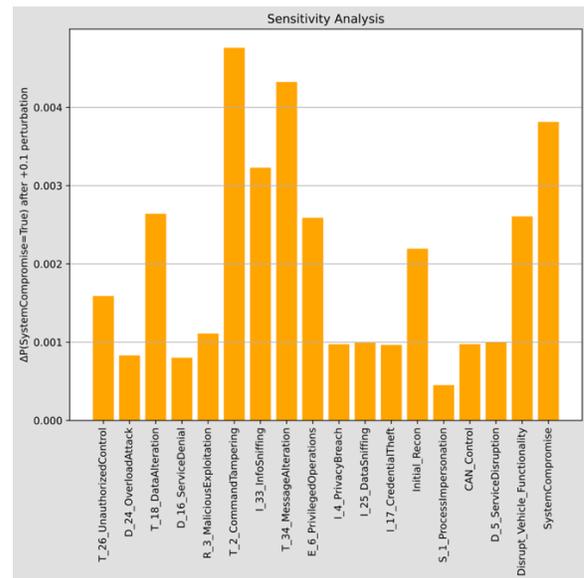

The sensitivity analysis results are shown in Figure 4 (bar chart), which displays the change in P(SystemCompromise=True) when each node's probability was perturbed by +0.1. A higher bar indicates greater influence of that node on the overall security outcome.

Notable observations include:

- The nodes T_OBC_2_CommandTampering, I_CBtoOBC_33_InfoSniffing, and T_CBtoOBC_34_MessageAlteration exhibited the highest sensitivity scores among root nodes, implying that even modest increases in their attack likelihood significantly raise the probability of system compromise.
- Intermediate nodes such as SystemCompromise, Disrupt_Vehicle_Functionality, and Initial_Recon also showed high influence, reflecting their centrality and cumulative impact in the attack propagation path.
- Nodes like S_OBC_1_ProcessImpersonation and D_OBC_5_ServiceDisruption had comparatively lower influence, suggesting a more limited role in the critical propagation chains under the current BN configuration.

This analysis provides actionable insights into which components are most critical to secure in order to reduce systemic risk. It also highlights nodes that, despite being lower in individual attack likelihood, may lie along high-impact propagation pathways.

# 8. Limitations

While BNs provide a structured and probabilistic approach for modeling cybersecurity risk, several limitations arise in both theory and practical implementation—particularly in the context of safety-critical systems like In-Vehicle Infotainment (IVI). These limitations are summarized below and should guide future improvements to the model presented in this paper.

## 8.1 Scalability and Complexity

As the number of threats and dependencies increases, the CPTs in BNs grow exponentially, making inference computationally expensive. This hinders the scalability of the model, particularly for systems with hundreds of interdependent components.

## 8.2 Dependence on Accurate Prior Probabilities

BNs require well-calibrated prior probabilities and conditional relationships to yield meaningful inferences. In this study, DREAD and SAHARA scores were used to estimate these values; however, such expert-derived metrics are inherently subjective and may not generalize across different threat contexts or systems.

## 8.3 Observability Assumptions

Many BN-based models—including this one—assume that threat states are fully observable or can be inferred with reliable certainty. In practice, cybersecurity monitoring is often partial, delayed, or noisy, which can distort posterior estimations and impact decision-making accuracy.

## 8.4 Static Structure and Causal Oversimplification

The causal relationships in BNs are pre-defined and static. They do not adapt to newly discovered attack vectors or evolving system architectures unless manually updated. Moreover, BNs may oversimplify complex causal relationships that are time-dependent or adversarial in nature.

## 8.5 Real-Time Constraints

Executing BN inference in real-time embedded environments (e.g., automotive systems) may introduce latency, especially when multiple intermediate nodes are used for layered propagation. Performance tuning is necessary to deploy BNs in safety-critical domains without compromising system responsiveness.

## 8.6 Limited Temporal Reasoning

This model captures probabilistic dependencies at a single point in time. It does not include temporal dynamics such as attacker dwell time, threat evolution, or sequence of compromise events—features that Dynamic Bayesian Networks (DBNs) or Hidden Markov Models (HMMs) could capture more effectively.

# 9. Conclusion

As cybersecurity threats grow increasingly dynamic, interdependent, and opaque, traditional static risk modeling approaches—such as attack trees, risk registers, or CVSS-based scoring—struggle to capture the real-world complexities of system compromise. This paper demonstrated that BNs provide a flexible and mathematically grounded alternative, enabling probabilistic inference, dynamic threat propagation modeling, and evidence-driven risk assessment.

By applying a structured model-to-model transformation process to a STRIDE-based threat analysis of an In-Vehicle Infotainment (IVI) system, this study showcased how BNs can effectively represent cascading cyber risks under uncertainty. The BN framework successfully quantified both direct and indirect escalation paths, revealed hidden interdependencies, and enabled dynamic updating of system compromise likelihoods based on observed threat indicators.

However, several limitations were also highlighted, including challenges related to scalability, dependence on expert-derived CPTs, static network structures, and limited support for temporal threat evolution. These findings point toward promising future directions, including the use of Dynamic Bayesian Networks (DBNs), automated structure learning algorithms, online parameter updating with real-time telemetry, and hybrid BN–machine learning models for adaptive cybersecurity defenses.

In conclusion, Bayesian Networks represent a valuable tool for modern cybersecurity operations—particularly when thoughtfully integrated with domain knowledge, continuous monitoring frameworks, and evolving threat intelligence. By advancing probabilistic modeling techniques, cybersecurity practitioners and researchers can move toward more resilient, transparent, and adaptive defense architectures in the face of growing cyber-physical threats.